\documentclass[11pt]{amsart}
\usepackage{geometry}                
\usepackage{graphicx}
\usepackage{amssymb}
\usepackage{epstopdf}
\title{Simple series solutions to specific heat-phonon spectrum inversion}
\author{Rong Qiang Wei}
\address{College of Earth and Planet Sciences, University of Chinese Academy of Sciences, Beijing, PRC, 100049}
\email{wrq1973@ucas.edu.cn}
\date{}

\begin{document}
\maketitle

{\bf Abstract} The specific heat-phonon spectrum inversion has played a significant role in solid physics. But for this inherently ill-posed problem, most of the known solutions are complex both in form and content, although they are rigorous and perfect. Here we suggest another simpler series solution to this problem, which can be easily calculated if the ratio of specific heat to temperature can be expanded into a power series, or specific heat can be expanded asymptotically and conditionally. Furthermore, we suggest similar solutions to the black-body radiation inversion.

\section{Introduction}

Phonon spectrum is a key in the study of various thermo-dynamical properties of the crystal. Once it is known, all the thermodynamic quantities related to the temperature can be obtained by integration. Phonon spectrum can be obtained inversely from specific heat, besides experimental determining and theoretical calculating. It is well known this inversion problem was first approximately solved by Einstein in 1907 and Debye in 1912, and led to the now famous Einstein model and Debye model, respectively.

Montroll (1942) and  Lifshitz (1954) independently reconsidered to invert phonon spectrum from the experimental heat capacity. Since then, various solutions to this problem have been provided. Some of them are analytical ones in integral form (Montroll 1942; Lifshitz, 1954; Kroll, 1952; Dai et al, 1990), or those in series (Chen, 1990), or asymptotic ones in series (Weiss, 1959; Loram, 1986). Others are numerical ones based on maximum entropy (Hague, 2005), or in Tikhonov regularization framework (Costa et al., 2014), or with some variational iteration approaches (Naumov and Musikhin, 2015; Musikhin et al.,2020). Since the numerical methods are relatively procedural, they will not be taken into account here. 

As will be seen in section 2, most analytical solutions above are related to integral (perhaps double integral ) of special functions such as Riemman Zeta function or Gamma function (eg. Dai et al., 1990), or to inverse Laplace transform (eg., Chen 1990). These may result in a complicated solution even for a simple function of specific heat, and requires special knowledge to get the final solution.

Here we will suggest another series solutions to the inversion problem of phonon spectrum from the specific heat. They are simpler, and require a less amount of labor to be evaluated.

\section{Simpler series inversion solutions to phonon spectrum}

If the phonon spectrum is known, the specific heat of lattice vibrations (forward problem) can be expressed as,

\begin{equation}\label{eq1}
C_v\left(T\right)=3R\int_{0}^{\infty}{\left(\frac{h\omega}{k_{_{\rm B}}T}\right)^2\frac{e^{h\omega/k_{_{\rm B}}T}}{\left(e^{h\omega/k_BT}-1\right)^2}}g\left(\omega\right){\rm d}\omega
\end{equation}
where $C_v(T)$ is the specific heat at constant volume, $T$ the temperature, $R$ the gas constant, $h$ the Planck's constant, $\omega$ the frequency,  and $k_{_{\rm B}}$ the Boltzmann constant, and $g\left(\omega\right)$ phonon spectrum.

The inversion problem is to recover the $g\left(\omega\right)$ based on the experimentally
measurable function of $C_v(T)$. 

\subsection{Previous studies}

\ \ \ \ 

Montroll (1942) obtained a solution to this inversion problem as,

\begin{equation}\label{eq2}
g\left( \omega \right) =\frac{1}{2\pi}\int_{-\infty}^{\infty}{\frac{\mathrm{d}u}{\zeta \left( 2+iu \right) \varGamma \left( 3+iu \right)}}\int_0^{\infty}{C\left( \theta \right)}\left( \omega \theta \right) ^{iu}\mathrm{d}\theta
\end{equation}

and $$C\left( \theta \right) =\int_0^{\theta}{\frac{\theta ^2\omega ^2g\left( \omega \right) \mathrm{d}\omega}{4\sinh ^2\left( \theta \omega /2 \right)}}$$

where $\theta=h/k_{_{\rm B}}T$, $C(\theta)=C_v(T)/k_{_{\rm B}}$.

The Montroll-Lifshitz formula (reviewed in Chen \& Rong (1998)) is,

\begin{equation}\label{eq3}
g\left( \omega \right) =\frac{1}{\omega}\int_{-\infty}^{\infty}{\frac{\left( h\omega \right) ^{ik\omega}\hat{C}\left( k \right) \mathrm{d}k}{\left( 1+\delta +ik \right) \zeta \left( 1+\delta +ik \right) \varGamma \left( 1+\delta +ik \right)}}
\end{equation}
where $\hat{C}\left( k \right)$ is the Fourier transform of $C_v(T)$. $\delta$ is chosen such that $0\leq\delta\leq 3$, in order to satisfy $\lim_{u\rightarrow\pm\infty}\Phi(u)=0$, and

$$
\varPhi \left( u \right) =\frac{e^{-\left( \delta +2 \right) u}e^{e^{-u}}}{\left( e^{e^{-u}}-1 \right) ^2}
$$

A solution in Grayson-Smith \& Stanley (1950) is,

\begin{equation}\label{eq4}
t^2f(t)=(2p/\pi)\sum_{m=1}^\infty a_m(1-\cos mt)/m
\end{equation}

$a_m$ is determined by, $$\gamma(\tau)=3R\sum_{m=1}^\infty a_m(\coth m\pi\tau -1/m\pi\tau)$$
where $pk_{\rm B}T/h=\tau$, $p\omega=t$, $C_v(T)=\gamma(\tau)$, $g(\omega)=f(t)$, $p$ is an arbitrary constant.

Kroll (1952) showed, 
 
 \begin{equation}\label{eq5}
 f\left( z \right) =\frac{2}{\pi}\int_0^{\infty}{\gamma \left( zt \right)}\mathrm{d}t\int_0^{\infty}{F\left( u \right) \cos tu\mathrm{d}u}
 \end{equation}
 
 and $$\gamma \left( z \right) =\sum_{\nu =0}^{\infty}{\left( -1 \right) ^{\nu}\frac{z^{2\nu}}{\left( 2\nu \right) !\left( 2\nu +4 \right) !\zeta \left( 2\nu +2 \right)}}$$
  
 where $C_v(x)=k_{_{\rm B}}x^3F(x)$, and finally $g(z)=z^2f(z)$.
 
A low-temperature asymptotic expansion by Weiss (1959) is,

\begin{equation}\label{eq6}
g\left( \omega \right) \sim \frac{1}{k_{_{\rm B}}}\sum_{n=0}^N{\frac{1}{\pi ^{2n+1}B_{n+1}}}\left( \frac{h}{2k_{_{\rm B}}} \right) ^{2n+1}\frac{C_{v}^{\left( 2n+1 \right)}\left( 0 \right)}{\left( 2n+1 \right) !}\omega ^{2n}
\end{equation}
where $C_v^{(n)}(0)$ denotes the $n'$th derivative of $C_v(T)$ evaluated at $T=0$. $B_n$ is the $n$'th Bernouilli number.

Similarly to Montroll-Lifshitz formula, Dai et al. (1990) gave a solution as,

\begin{equation}\label{eq7}
g\left( \omega \right) =\frac{1}{\omega}\int_{-\infty}^{\infty}{\frac{\left( h\omega /T_0 \right) ^{ik-s}\tilde{Q}\left( k \right) \mathrm{d}k}{\left( 1+s+ik \right) \varGamma \left( 1+s+ik \right) \zeta \left( 1+s+ik \right)}}
\end{equation}
where $Q(x)=C_v(T_0e^x)$, $\tilde{Q}$ is the Fourier transform. $s$ is a parameter.

Based on the M$\ddot{\rm o}$bius inversion technique, Chen (1990) provided a solution as,

\begin{equation}\label{eq8}
g\left( \omega \right) =\frac{1}{rk_{_{\rm B}}\omega ^2}\sum_{n=1}^{\infty}{\mu \left( n \right)}L^{-1}\left[ \frac{C_v\left( h/nk_{_{\rm B}}u \right)}{nu^2}; u\rightarrow \omega \right]
\end{equation}
where $L^{-1}[\  ] $ represents the inverse Laplace transform. $\mu (n)$ is M$\ddot{\rm o}$bius function. $r$ is the number of atoms per unit cell.

Hughes et al. (1990) re-obtained (Lifshitz (1954) mentioned this formal expression of the solution based on Mellin transform),

\begin{equation}\label{eq8_add}
g\left( \omega \right) =M^{-1}\left[ \frac{\hat{C}\left( 1-s \right)}{\varGamma \left( 3-s \right) \zeta \left( 2-s \right)} \right] 
\end{equation}
where $M^{-1}[\  ]$ is the inverse Mellin transform, $\hat{C}$ the Mellin transform of $C_v(T)$.

These previous studies are very helpful to understand the specific heat-phonon spectrum inversion problem. These solutions are rigorous and perfect. Based on them, many further studies have been carried out (eg., Wen et al., 2003; Richet, 2009; Ji et al., 2012).

\subsection{A simple series inversion solution to inversion problem}

Comparing Eq. (\ref{eq2}) to Eq. (\ref{eq8_add}), it can be seen that Eq. (\ref{eq4}) and Eq. (\ref{eq6}) are relative simple, while Eq. (\ref{eq2}), Eq. (\ref{eq3}) ,Eq. (\ref{eq5}), Eq. (\ref{eq7}) - Eq. (\ref{eq8_add}) are complicated.  Because the latter 6 solutions include integral in complex domain of the special functions and/or Fourier transform, or inverse Laplace transform, or Mellin transform, all of them demand a large amount of labor. 

In this subsection, we provide another solution to the specific heat-phonon spectrum inversion. We start from Eq. (\ref{eq9}) in the following,

\begin{equation}\label{eq9}
f\left( x \right)=\int_0^{\infty}{K\left( \frac{y}{x} \right) g\left( y \right) {\rm d}y}
\end{equation}
where $K(\ )$ denotes integral kernel function. 

Let $y=xt$ and assume $g(y)=\sum_ng_ny^n$,

\begin{equation}\label{eq10}
\begin{aligned}
f\left( x \right)&		=\int_0^{\infty}{K\left( \frac{y}{x} \right) g\left( y \right) {\rm d}y}\\
  	&		=\int_0^{\infty}{K\left( t \right) g\left( xt \right)x {\rm d}t}\\
	&		=\int_0^{\infty}{K\left( t \right)}\left(\sum_n g_n x^n t^n\right)x{\rm d}t\\
	&		=\sum_n g_n x^{n+1}\int_0^{\infty}{K\left( t \right)}t^n {\rm d}t\\
	&		=\sum_n \bar{g_n} x^{n+1}\\
\end{aligned}
\end{equation}
where $\bar{g_n}=g_n \int_0^{\infty}{K\left( t \right)}t^n {\rm d}t$.

If $f(x)=\sum_n a_n x^{n+1}$ or $\frac{f(x)}{x}=\sum_n a_n x^n$ and by comparing the coefficients on both sides of Eq. (\ref{eq10}), we can obtain $\bar{g_n}$, then $g_n$, and then  $g(y)=\sum_ng_ny^n$. 

{\bf Remark 1:} In Eq. (\ref{eq10}), $n$ may be replaced by any real or complex $\gamma$ for which the integral $\int_0^{\infty}{K\left( t \right)}t^\gamma {\rm d}t$ has a meaning.

{\bf Remark 2:} The corresponding homogeneous equation to Eq. (\ref{eq10}) is $h\left( x \right)=\lambda\int_0^{\infty}{K\left( t \right) h\left( xt \right) {\rm d}t}$. It has a special solution $f\left(x\right)=x^\gamma$ where $\gamma$ is any value for which the integral has a meaning.

{\bf Remark 3:} The upper bound of $\sum_n$ may be finite or infinite, depending on the expansion of $f(x)$.

{\bf Remark 4:} The approach from Eq. (\ref{eq10}) can be extended to any integral equation of $f\left( x \right)=\int_0^{\infty}{K\left( x^\alpha y^\beta \right) g\left( y \right) {\rm d}y}$, where $\alpha$, $\beta$ are real.

Similarly, if $g(y)=\sum_n{g_n}/{y^n}$ then,

\begin{equation}\label{eq10_add}
\begin{aligned}
f\left( x \right)&		=\int_0^{\infty}{K\left( t \right)}\left(\sum_n \frac{g_n}{x^n t^n}\right)x{\rm d}t\\
  	&		=\sum_n \frac{g_n} {x^{n-1}}\int_0^{\infty}{K\left( t \right)}t^{-n} {\rm d}t\\
	&		=\sum_n \frac{\bar{g_n}} {x^{n-1}}\\
\end{aligned}
\end{equation}
where $\bar{g_n}=g_n \int_0^{\infty}{K\left( t \right)}t^{-n} {\rm d}t$. Similarly above, we can obtain the explicit form of $g(y)=\sum_n g_n/y^n$, if $f(x)=\sum_n {a_n}/{x^{n}}$. 

\ \ \ 

We return to Eq. (\ref{eq1}) with the approach from Eq. (\ref{eq9})-Eq. (\ref{eq10}). Let $z=h\omega/k_{_{\rm B}}T$, $\omega=k_{_{\rm B}}Tz/h$, one can get,

\begin{equation}\label{eq11}
\frac{C_v\left(T\right)}{T}=3R\frac{k_{_{\rm B}}}{h}\int_{0}^{\infty}\frac{z^2e^z}{\left(e^z-1\right)^2}g\left(\frac{k_{_{\rm B}}T}{h}z\right){\rm d}z
\end{equation}

Substituting that $C_v\left(T\right)/T=\sum_n{a_nT^n}$, $g\left(\omega\right)=\sum_n {g_n\omega^n}$ into Eq. (\ref{eq11}), and by comparing the coefficients on both sides, we have,

\begin{equation}\label{eq12}
g_n=\frac{a_nB}{\int_{0}^{\infty}\frac{z^{n+2}e^z}{\left(e^z-1\right)^2}{d}z}\left(\frac{h}{k_{_{\rm B}}}\right)^n=\frac{a_nB}{\varGamma\left(n+3\right)\zeta\left(n+2\right)}\left(\frac{h}{k_{_{\rm B}}}\right)^n
\end{equation}
where $B=h/3Rk_{_{\rm B}}$. 

Finally we obtain,

\begin{equation}\label{eq13}
g(\omega)=\sum_n g_n\omega^n=\sum_n \left[\frac{a_nB}{\varGamma\left(n+3\right)\zeta\left(n+2\right)}\left(\frac{h}{k_{_{\rm B}}}\right)^n\right]\omega^n
\end{equation}

Similarly, if $C_v\left(T\right)=\sum_na_n/T^n$ and with Eq. (\ref{eq10_add}), then

\begin{equation}\label{eq14}
g(\omega)=\sum_n \frac{g_n}{\omega^n}=\sum_n \left[\frac{a_n B}{\varUpsilon (2-n)}\left(\frac{h}{k_{_{\rm B}}}\right)^{-n}\right]\omega^{-n}
\end{equation}
where $\varUpsilon (2-n)$ can be obtained from the analytic continuation of $\int_{0}^{\infty}\frac{z^{n+2}e^z}{\left(e^z-1\right)^2}{\rm d}z$, if $2-n<0$. It should be pointed out that $n$ in $C_v\left(T\right)=\sum_na_n/T^n$ should make $\varUpsilon (2-n)\neq 0$ or $\infty$.

If $C_v\left(T\right)/T=\sum_n{a_nT^n}$, $g(\omega)$ can be inferred from Eq. (\ref{eq13}). It can be seen this series only involves Gamma function and Riemann zeta function with integer arguments; It is easier to be calculated than most of the solutions mentioned in subsection 2.1. If $C_v\left(T\right)=\sum_na_n/T^n$, $g(\omega)$ can be inferred from Eq. (\ref{eq14}). It can also be seen that this asymptotic series is easier to be evaluated, although it is a conditional solution.

\subsection{Applications}

\ \ \ \ 

(1) $C_v(T)$ obtained at low-temperatures

At this time, $T\rightarrow 0$, and 

\begin{equation}\label{eq15}
C_v(T)=a_3T^3+a_5T^5+\ldots=\sum_{n=1}a_{_{2n+1}}T^{2n+1}
\end{equation}

From Eq. (\ref{eq15}) and Eq. (\ref{eq13}), we can get,

\begin{equation}\label{eq16}
g(\omega)=\sum_n \frac{a_{_{2n+1}}B}{\varGamma\left(2n+3\right)\zeta\left(2n+2\right)}\left(\frac{h}{k_{_{\rm B}}}\right)^{2n}\omega^{2n}
\end{equation}

Eq. (\ref{eq16}) is very similar to Weiss (1959) and Chen \& Rong (1998).

\ \ \ \

(2) $C_v(T)$ obtained at high-temperatures

At this time, $T\rightarrow \infty$, and 

\begin{equation}\label{eq17}
C_v(T)=b_0-\frac{b_2}{T^2}+\frac{b_4}{T^4}+\ldots=\sum_{n=1}(-1)^{n+1} \frac{b_{_{2n-2}}}{T^{2n-2}}
\end{equation}

From Eq. (\ref{eq17}) and Eq. (\ref{eq14}), we can get in form,

\begin{equation}\label{eq18}
g(\omega)=\sum_n \left[\frac{(-1)^{n+1}b_{_{2n-2}}B}{\varUpsilon (4-2n)}\left(\frac{h}{k_{_{\rm B}}}\right)^{2-2n}\right]\frac{1}{\omega^{2n-2}}
\end{equation}

If $\omega\rightarrow\infty$, and let $n=1$, $g(\omega)\sim  \frac{b_0B}{\varUpsilon (2)}=\frac{b_0B}{\varGamma(3)\zeta (2)}$ from Eq. (\ref{eq18}). This means $g(\omega)$ is a constant when $T\rightarrow\infty$, which is consistent with  
Einstein model.

\ \ \ \ 

(3) A numerical example

\begin{figure}[htb]
\setlength{\belowcaptionskip}{0pt}
\centering
\includegraphics[scale=0.55]{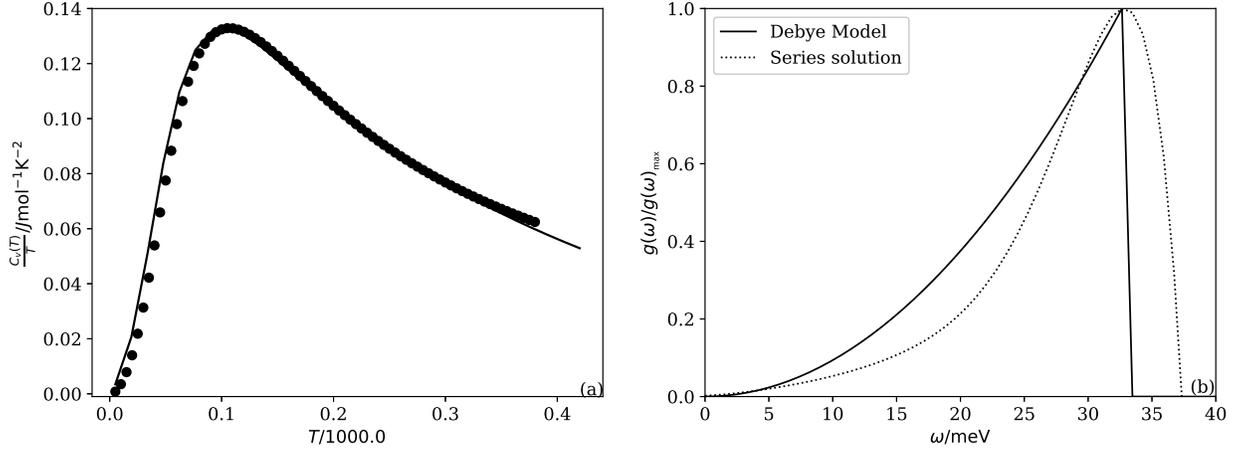}
\renewcommand{\figurename}{Fig.}
\caption{(a) $C_v(T)/T$ of the Cu varies with normolized temperatures, in which $C_v(T)$ are calculated with Debye Model ($\Theta_D=382 K$ (Marder, 2010)). The solid line is the approximate curve. (b) Phonon spectrum $g(\omega)$ from the solution of Eq. (\ref{eq13}) ($n=10$) and Debye model. Each curve of $g(\omega)$ is normalized by its maximums.}
\label{fig1}
\end{figure}

In most cases, we can only obtain some discrete $C_v(T)$ from experiments. Here we apply Eq. (\ref{eq13}) to Cu to inverse its phonon spectrum $g(\omega)$. Firstly the $C_v(T)$ of Cu is calculated by Debye Model with $\Theta_D=382 K$ (Marder, 2010) and then $C_v(T)/T$ is calculated (Dots in Fig. \ref{fig1}a). To get the power expansion of these discrete $C_v(T)/T$, a model of $\sum_{i=1}^{6}{c_i\exp\left(s_iT/T_m\right)}$ (where $c_i,\ s_i$ are parameters. $T_m$ is the temperature used for normalization) is secondly used to approximate $C_v(T)/T$ (Solid line Fig. \ref{fig1}(a)). Thirdly this model is expanded into a power series to get $a_n$ and further to obtain $g_n$ with Eq. (\ref{eq12}). And finally the normalized $g(\omega)$ is obtained from Eq. (\ref{eq13}) ($n=10$) and shown in Fig. \ref{fig1}(b).

It can be found from Fig. \ref{fig1} that the main properties of $g(\omega)$ from Eq. (\ref{eq13}) are consistent with those from the Debye model. These properties include the curve shape and cut-off frequency. However, such a good result is from a good approximation of $C_v\left(T\right)/T$. In other word, this solution in series here depends on how to get a smooth $C_v\left(T\right)/T$. 

\section{Series solution to the black-body radiation inversion}

We also can use the method in section 2 to the black-body radiation inversion.

\begin{equation}\label{eq19}
W\left( \omega \right) =\frac{2h\omega ^3}{c^2}\int_0^{\infty}{\frac{a\left( T \right)}{e^{h\omega /k_{\mathrm{B}}T}-1}}\mathrm{d}T
\end{equation}
where $W\left( \omega \right)$ is the power spectrum, $c$ the speed of light, $a(T)$ the area-temperature distribution on the surface.

Let $z=h\omega/k_{_{\rm B}}T$, we can obtain,

\begin{equation}\label{eq20}
W\left( \omega \right) =\frac{2h\omega ^3}{c^2}\int_0^{\infty}{\frac{z^{-2}a\left( h\omega /k_{\mathrm{B}}z \right)}{e^z-1}\left( \frac{h\omega}{k_{\mathrm{B}}} \right)}\mathrm{d}z
\end{equation}

If $a(T)=\sum_n{a_nT^n}$, we can obtain in form,

\begin{equation}\label{eq21}
\frac{W\left( \omega \right)}{\omega ^4}=\sum_n{\left[ a_n\frac{2h}{c^2}\left( \frac{h}{k_{\mathrm{B}}} \right) ^{n+1}\int_0^{\infty}{\frac{z^{-2-n}}{e^z-1}}\mathrm{d}z \right] \omega ^n}
\end{equation}

Then, if $W(\omega)/\omega^4=\sum_n{w_{2n}\omega^{2n}}$, 

\begin{equation}\label{eq22}
\begin{aligned}
a(T)=\sum_n {a_{2n}T^{2n}}&=\sum_n \left[\frac{w_{2n}}{\frac{2h}{c^2}\left( \frac{h}{k_{\mathrm{B}}} \right) ^{2n+1}\zeta(-1-2n)}\right]T^{2n}
\end{aligned}
\end{equation}

If $a(T)=\sum_na_n/T^n$, we can obtain in form,

\begin{equation}\label{eq23}
\frac{W\left( \omega \right)}{\omega ^4}=\sum_n{\left[ a_n\frac{2h}{c^2}\left( \frac{h}{k_{\mathrm{B}}} \right) ^{1-n}\int_0^{\infty}{\frac{z^{n-2}}{e^z-1}}\mathrm{d}z \right] \omega ^{-n}}
\end{equation}

Then if $W(\omega)/\omega^4=\sum_n w_n/\omega^n$ where $n\neq 2$,

\begin{equation}\label{eq24}
\begin{aligned}
a(T)=\sum_n \frac{a_n}{T^n}&=\sum_n \left[\frac{w_n}{\frac{2h}{c^2}\left( \frac{h}{k_{\mathrm{B}}} \right) ^{1-n}\zeta(n-1)}\right]T^{-n}
\end{aligned}
\end{equation}

\section{Conclusion}

Simpler series solutions to the specific heat-phonon spectrum inversion are suggested, which can be more easily evaluated if the specific heat is a smooth function, or specific heat can be asymptotically and conditionally. These series solutions may be helpful to discuss some related problems of solid physics.

\vspace{5em}


\ \

\end{document}